\documentclass[
    ,final            
  ]
  {aipproc}

\layoutstyle{6x9}

\begin{document}

\title{Observational Signatures of the First Galaxies}


\author{Jarrett L. Johnson}{
  address={Max-Planck-Institut f{\"u}r extraterrestrische Physik, 
Giessenbachstra\ss{}e, 85748 Garching, Germany \\
Theoretical Modeling of Cosmic Structures Group \\}
}

\begin{abstract}
Detection of the radiation emitted from some of the earliest galaxies will be made possible in the next decade, with the launch of the 
James Webb Space Telescope (JWST).  A significant fraction of these galaxies may host Population (Pop) III star clusters.  The detection 
of the recombination radiation emitted by such clusters would provide an important new constraint on the initial mass function (IMF) of 
primordial stars.  Here I review the expected recombination line signature of Pop III stars, and present the results of cosmological 
radiation hydrodynamics simulations of the initial stages of Pop III starbursts in a first galaxy at $z$ $\sim$ 12, from which the 
time-dependent luminosities and equivalent widths of IMF-sensitive recombination lines are calculated. While it 
may be unfeasible to detect the emission from Pop III star clusters in the first galaxies at $z$ > 10, even with next generation 
telescopes, Pop III star clusters which form at lower redshifts (i.e. at $z$ < 6) may be detectable in deep surveys by the JWST.  

\end{abstract}

\maketitle


\section{Introduction}
In this contribution, I will address three key questions pertaining to the observational signatures of the first galaxies,
and in particular to the prospects for the detection and identification of Pop III stars.  These questions are the following: 
\begin{itemize}

\item
How can Population (Pop) III stars be identified observationally and their initial mass function (IMF) constrained?
\item 
How long did Pop III star formation continue after the epoch of the first stars?         
\item  
Will observational facilities in the coming years, and in particular the {\it James Webb Space Telescope} (JWST), 
be able to detect and identify Pop III stellar populations?

\end{itemize}
The next three sections are devoted to addressing these three questions in the order given above; in the final section I will briefly summarize
the main conclusions.

\begin{figure}
  \includegraphics[height=.3\textheight]{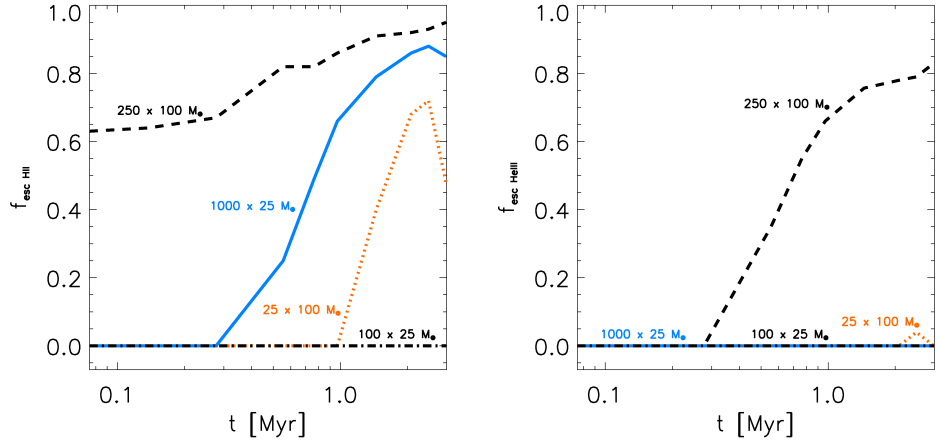}
  \caption{The escape fraction of H~{\sc i}-ionizing photons ({\it left panel}) and that of He~{\sc ii}-ionizing photons ({\it right panel}), from a Pop III star cluster formed in a first galaxy at $z$ $\sim$ 12.  Each line corresponds to a different choice of IMF and total number of stars, as labeled.  There is a tight anticorrelation between the escape fraction of H~{\sc i}-ionizing photons and the emission in hydrogen recombination lines (e.g. Ly$\alpha$ and H$\alpha$; see Figs. 2 and 3).  For most cases, however, the negligible escape fraction of He~{\sc ii}-ionizing photons leads to a tight correlation between the luminosity emitted in the 
He~{\sc ii} $\lambda$1640 line and the total mass contained in stars, for a given IMF (see \protect\cite{JLJetal2009}).}
\end{figure}

\section{The spectral signature of Pop III star formation}
It is well-known that Pop III stars are likely to be considerably hotter than present-day stars, for a given stellar mass (e.g. \cite{bond, cast, eleid, ezer}).  The high surface temperatures 
of Pop~III stars lead to enhanced emission of ionizing radiation, in particular photons with energies above 54.4 eV which can ionize He~{\sc ii}.  This, in turn, implies that 
the photoionized regions surrounding Pop~III stars should shine brightly in He~{\sc ii} recombination lines, principally in He~{\sc ii} $\lambda$1640 \cite{bkl,oh,tum,S02}.
Indeed, a number of observational efforts have been geared toward detecting this recombination line as an indicator of Pop~III star formation \cite{n2005,n2008} (see also \cite{bou, daw, fos, pres, shap, wang}), although to date no definitive detections on Pop III star formation have been reported.  Nonetheless, the He~{\sc ii} $\lambda$1640 signature is likely one of the most reliable indicators of metal-free star formation and will continue to be sought using future telescopes (see e.g. \cite{barton}).

Helium recombination lines also afford a means to constrain the IMF of Pop III stars, potentially allowing to test different theoretical predictions for the characteristic mass of Pop III stars 
(e.g. of the order of 10 M$_{\odot}$ or 100 M$_{\odot}$).  However, constraining the IMF using the ratio of the observed fluxes in He~{\sc ii} $\lambda$1640 and H$\alpha$ or Ly$\alpha$ 
poses some challenges, owing to the evolution of the flux emitted in these lines.  While the evolution of the massive stars in a cluster will alter its spectral characteristics \cite{S02,S03}, 
the photoheating of the gas surrounding the cluster will also lead to similar evolution \cite{JLJetal2009}.  
In particular, as the gas surrounding the cluster is photoheated it expands, thereby allowing for the escape of ionizing photons into the intergalactic 
medium (IGM); as more ionizing photons escape, fewer are available to ionize the dense gas from which recombination lines are emitted, and the luminosity in those lines correspondingly drops.    

\begin{figure}
  \includegraphics[height=.3\textheight]{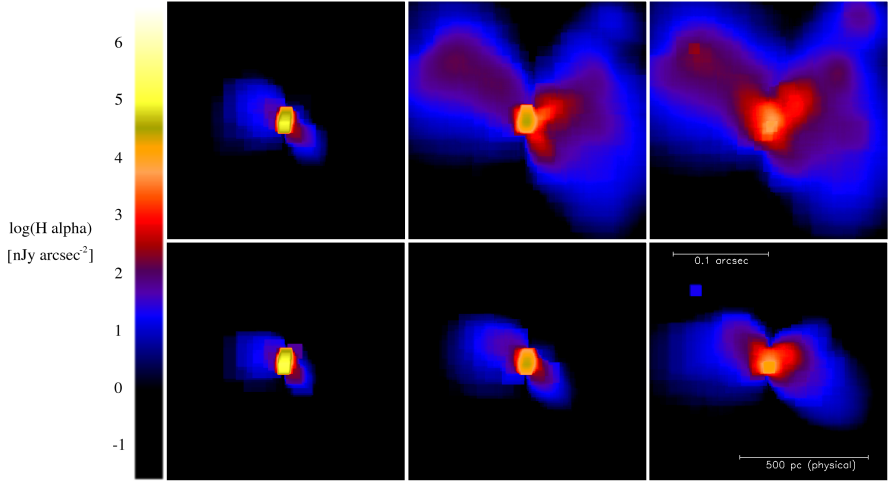}
  \caption{The flux in H$\alpha$ per square arcsecond, emitted from Pop III stellar clusters at $z$ $\sim$ 12, as observed on the sky at $z$ = 0, 
assuming a spectroscopic resolution of R = 1000.  Shown here are the two most massive of the four simulated stellar clusters presented in \protect\cite{JLJetal2009}, one containing 
25 ${\rm M}_{\odot}$ stars ({\it bottom panels}), the other containing 100 ${\rm M}_{\odot}$ stars ({\it top panels}).  From left to right, the 
clusters are shown at 10$^5$ yr, 1 Myr, and 3 Myr after formation.   
The highest total fluxes occur at the earliest times, before the H~{\sc ii} region has broken out of the galaxy; hence, the youngest stellar clusters are the 
most readily observed.}
\end{figure}

The escape fraction of ionizing photons from Pop III stellar clusters forming in a first galaxy at $z$ $\sim$ 12, as calculated from the simulations presented in \cite{JLJetal2009}, 
are shown in Figure 1.  As this Figure shows, the escape fractions
of H~{\sc i}- and He~{\sc ii}-ionizing photons can differ greatly, leading to evolution of the ratio of the luminosities emitted in H~{\sc i} and He~{\sc ii} recombination lines, thus
complicating the use of such ratios as indicators of the IMF.   For two simulations presented in \cite{JLJetal2009}, the evolution of the flux visible in H$\alpha$ 
is shown in Figure 2; as the escape fraction of H~{\sc i}-ionizing photons generally increases with time, the total flux is highest at early times, making the youngest clusters the most easily observed.     

Figure 3 shows the equivalent width (EW) of three prominent recombination lines for each of the four simulations presented in \cite{JLJetal2009}.  While the ratio of the fluxes in He~{\sc ii} and H~{\sc i} recombination lines can be a problematic indicator of the stellar IMF, this Figure shows that the EW of He~{\sc ii} $\lambda$1640 may be a more robust indicator, always being larger for clusters with the more top-heavy IMF regardless of the total mass in stars.

\begin{figure}
  \includegraphics[height=.3\textheight]{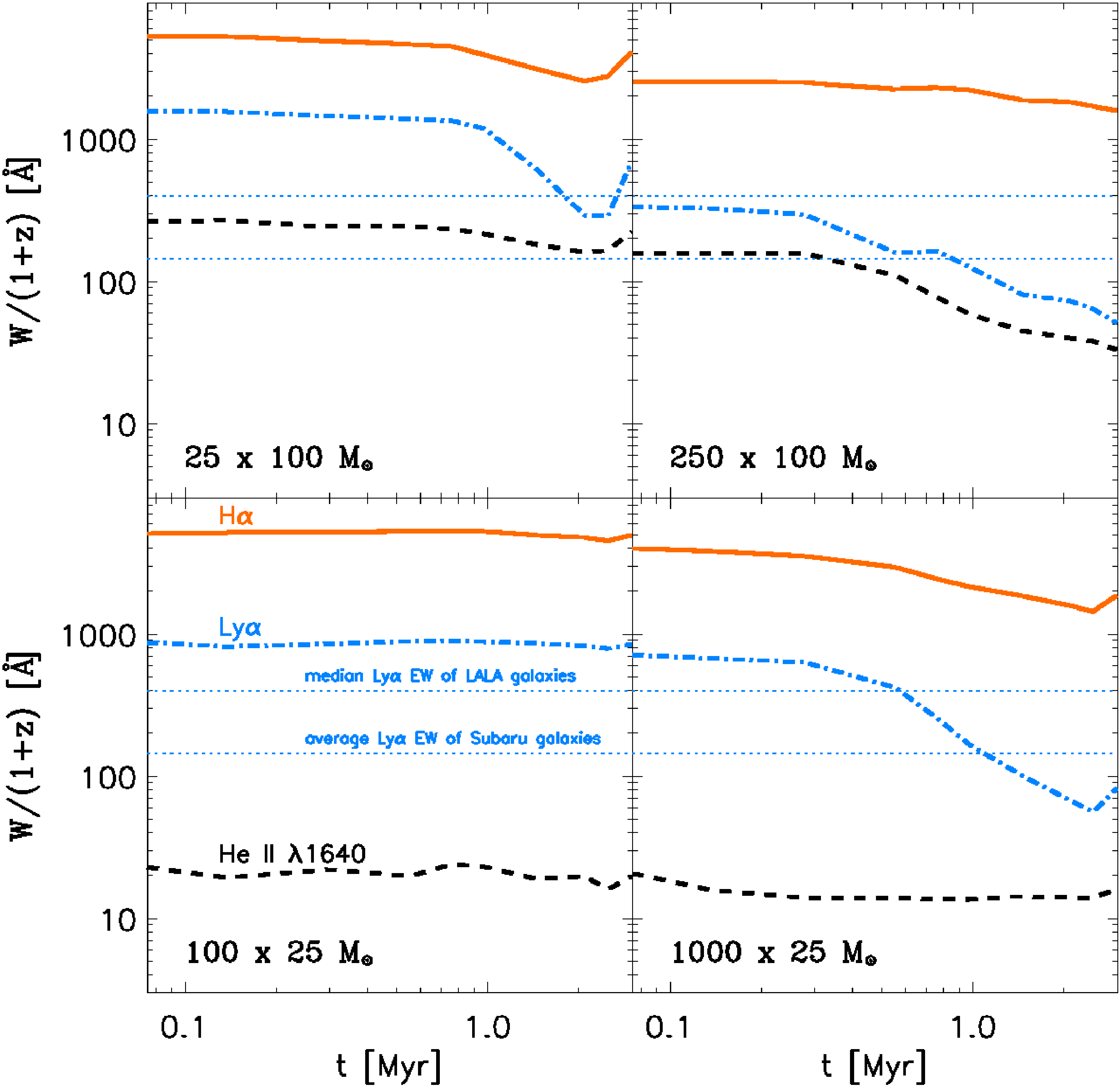}
  \caption{The rest frame EWs of Ly$\alpha$, H$\alpha$, and He~{\sc ii} $\lambda$1640, as a function of time, for the same clusters as shown in Figs. 1 and 2.
The {\it dotted} lines show the observed EWs of galaxies from two different surveys carried out at $z$ = 4.5 and $z$ $\ge$ 6 (\protect\cite{MR2002}, \protect\cite{N2007}).  The He~{\sc ii} 
$\lambda$1640 EW for clusters of 100 ${\rm M}_{\odot}$ stars is always higher than that for 25 ${\rm M}_{\odot}$ stars, regardless of the total stellar 
mass in the clusters; hence, we conclude that the EW of this line is a robust indicator of a very top-heavy IMF (see \protect\cite{JLJetal2009}).}
\end{figure}

\section{Pop III star clusters formed after reionization}
It is likely that even surveys to be carried out by the JWST will not be deep enough to detect the first stars or galaxies at $z$ $\ge$ 10 \cite{bark,gardner,JLJetal2009,ric}, although 
Pop III supernovae (e.g. \cite{hai,trenti,wein,wise}) and perhaps stars powered by dark matter (DM) annihilation \cite{freese, zack} may still be detected.   
This provides motivation to consider whether Pop III star clusters may form also at lower redshifts, where IMF-sensitive recombination lines may be more readily detected.

In regions of the universe that undergo reionization at sufficiently early times (i.e. $z$ $\sim$ 20), thereby quenching star formation in DM minihaloes, Pop III star clusters may form in the unenriched descendants of those minihalos at $z$ $\leq$ 6.  Figure 4 shows the predicted abundance of such clusters, for various assumptions on the reionization history, the minimum mass of halos which may host star formation after reionization, and the 
speed of external metal enrichment by neighboring galaxies \cite{JLJ2010}.  As this Figure shows, although likely to be very rare, such Pop III clusters may be abundant enough to be detected in the Deep-Wide
Survey (DWS) to be carried out by the JWST \cite{gardner, wind}.

\section{Prospects for detection in JWST deep surveys}
If the number density of Pop III stellar clusters at low redshift (e.g. $z$ $\leq$ 6) is indeed high enough for some of them to lie within the area to be surveyed by the JWST, the question 
remains whether these clusters would be bright enough to be detected.  Figure 5 shows the monochromatic Ly$\alpha$ flux predicted for such clusters, for different assumptions on the IMF, 
the minimum halo mass for star formation, and the star formation efficiency \cite{JLJ2010}.  As the Figure shows, for a very top-heavy IMF and/or a high star formation efficiency, the JWST 
NIRCam may detect these clusters in the planned DWS.  Furthermore, spectroscopic follow-up with NIRSpec may detect the He II $\lambda$1640 flux, 
thereby allowing both for the confirmation of candidate Pop III star clusters and for constraints to be placed on the stellar IMF \cite{JLJ2010} (see also \cite{dij}).

It should be noted that strong He II $\lambda$1640 emission, while a telltale sign of Pop III star formation, does not in itself 
prove the existence of Pop III stars. Other observational signatures of Pop III star formation should thus be pursued.  Among the other signs of 
Pop III star formation may be a distinct absence of metal emission or absorption lines.  Additionally, unless the mixing of metals with the primordial 
gas is sufficiently slow (see e.g. \cite{JH2006}), one would expect Pop III star clusters not to form in galaxies in which previous star formation and 
metal enrichment have occurred; instead, as discussed above, Pop III star formation may occur only in previously unenriched DM halos at $z$ $\leq$ 6, and furthermore 
perhaps only in halos within typical distances of $\sim$ 1 Mpc comoving of galaxies which begin reionizing the IGM at $z$ $\sim$ 20 \cite{JLJ2010}.


\section{Summary}
In closing, I would like to highlight the following conclusions corresponding to the three key questions addressed in the work presented here:
\begin{itemize}

\item
The Pop III IMF can be constrained with detection of helium recombination emission (particularly the He II  $\lambda$1640 line).  
This emission varies due to both stellar evolution and hydrodynamic evolution of photoionized regions.

\item  
In rare regions which are reionized at early times, Pop III star formation may extend well beyond the epoch of the first stars and galaxies.

\item 
If this is so, then planned JWST surveys may detect Pop III stellar clusters at redshifts $z$ < 6 and allow for constraints to be placed on the IMF.
         
\end{itemize}

\begin{figure}
  \includegraphics[height=.26\textheight]{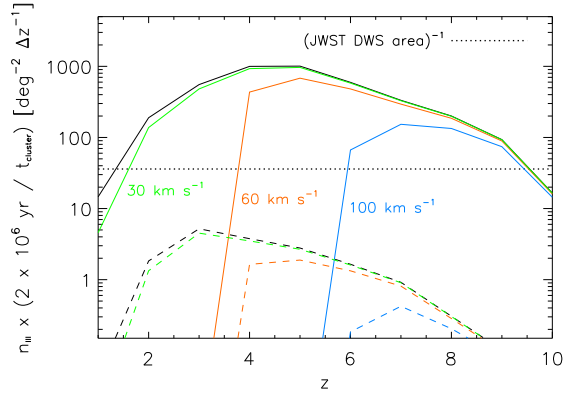}
  \caption{The number density $n_{\rm III}$ of Pop~III star clusters formed in reionized regions of the universe,  
taking into account enrichment of the IGM by galactic winds, and normalized to a cluster lifetime $t_{\rm cluster}$ = 2 Myr.  
The solid lines correspond to a minimum halo circular velocity for star formation 
of 20 km~s$^{-1}$, while the dashed lines correspond to a minimum of 30 km~s$^{-1}$.  For each series of lines, the top ({\it black}) line is 
a model which neglects external metal enrichment, while the colored lines correspond to different metal-enriched wind velocities, as labeled.    
The number density at which one cluster per unit redshift is expected to be within the planned JWST Deep-Wide Survey area is shown by the {\it dotted} line (see \protect\cite{JLJ2010}).}
\end{figure}


\begin{theacknowledgments}
I would like to thank the conference organizers, Dan Whalen, Naoki Yoshida, and Volker Bromm, for hosting a most enjoyable and productive event, 
as well as for allowing me to present this work.  I am also grateful for support from the Theoretical Modeling of Cosmic Structures (TMoX) Group at MPE.        

\end{theacknowledgments}



\bibliographystyle{aipproc}   

\bibliography{sample}

\begin{thebibliography}{99}

\bibitem{bark}Barkana R., Loeb A. 2000, ApJ, 531, 613
\bibitem{barton}Barton E.~J., Dav{\' e} R., Smith J.-D.~T., Papovich C., Hernquist L., Springel V. 2004, ApJ, 604, L1
\bibitem{bond}Bond J.~R., Arnett W.~D., Carr B.~J. 1984, ApJ, 280, 825
\bibitem{bou}Bouwens R.~J., et al. 2009, ApJ, submitted (arXiv:0910.0001)
\bibitem{bkl}Bromm V., Kudritzki R.~P., Loeb A. 2001, ApJ, 552, 464
\bibitem{cast}Castellani V., Chieffi A., Tornambe A. 1983, ApJ, 272, 249
\bibitem{daw}Dawson S., et al. 2004, ApJ, 617, 707
\bibitem{dij}Dijkstra M., Lidz A., Wyithe J.~S.~B. 2007, MNRAS, 377, 1175
\bibitem{eleid}El Eid M.~F., Fricke K.~J., Ober W.~W. 1983, A\&A, 119, 54
\bibitem{ezer}Ezer D. \& Cameron A.~G.~W. 1971, Ap\&SS, 14, 399
\bibitem{fos}Fosbury R.~A.~E., et al. 2003, ApJ, 596, 797
\bibitem{freese}Freese K., Ilie C., Spolyar D., Valluri M., Bodenheimer P. 2010, arXiv:1002.2233
\bibitem{gardner}Gardner J.~P., et al. 2006, SSRv, 123, 485
\bibitem{hai}Haiman Z. 2008, {\it Astrophysics in the Next Decade: JWST and Concurrent Facilities}, Ap\&SS Library, Eds. H. Thronson, A. Tielens, M. Stiavelli (arXiv:0809.3926)
\bibitem{JH2006}Jimenez R., Haiman Z. 2006, Nat, 440, 501
\bibitem{JLJ2010}Johnson J.~L. 2010, MNRAS, in press (arXiv:0911.1294)
\bibitem{JLJetal2009}Johnson J.~L., Greif T.~H., Bromm V., Klessen R.~S., Ippolito J. 2009, MNRAS, 399, 37
\bibitem{MR2002}Malhotra S., Rhoads J.~E. 2002, ApJ, 565, L71
\bibitem{n2005}Nagao T., et al. 2005, ApJ, 631, L5
\bibitem{N2007}Nagao T., et al. 2007, A\&A, 468, 877
\bibitem{n2008}Nagao T., et al. 2008, ApJ, 680, 100
\bibitem{oh}Oh P., Haiman Z., Rees M.~J. 2001, ApJ, 553, 73 
\bibitem{pres}Prescott M.~K.~M., Dey A., Jannuzi B.~T. 2009, ApJ, 702, 554
\bibitem{ric}Ricotti M., Gnedin N.~Y., Shull J.~M. 2008, ApJ, 685, 21
\bibitem{S02}Schaerer D. 2002, A\&A, 382, 28
\bibitem{S03}Schaerer D. 2003, A\&A, 397, 527
\bibitem{shap}Shapley A.~E., Steidel C.~C., Pettini M., Adelberger K.~L. 2003, ApJ, 588, 65
\bibitem{trenti}Trenti M., Stiavelli M., Shull J.~M. 2009, ApJ, 700, 1672
\bibitem{tum}Tumlinson J., Giroux M.~L., Shull J.~M. 2001, ApJ, 550, L1
\bibitem{wang}Wang J.~X., Malhotra S., Rhoads J.~E., Zhang H.~T., Finkelstein S.~L. 2009, ApJ, 706, 762
\bibitem{wein}Weinmann S.~M. \& Lilly S.~J. 2005, ApJ, 624, 526 
\bibitem{wind}Windhorst R. A., Cohen S. H., Jansen R. A., Conselice C., Yan H. 2006, New Astron. Rev., 50, 113
\bibitem{wise}Wise J.~H. \& Abel T. 2005, ApJ, 629, 615
\bibitem{zack}Zackrisson E., et al. 2010, ApJ, submitted (arXiv:1002.3368)

\end{thebibliography}

\IfFileExists{\jobname.bbl}{}
 {\typeout{}
  \typeout{******************************************}
  \typeout{** Please run "bibtex \jobname" to optain}
  \typeout{** the bibliography and then re-run LaTeX}
  \typeout{** twice to fix the references!}
  \typeout{******************************************}
  \typeout{}
 }

\begin{figure}
  \includegraphics[height=.23\textheight]{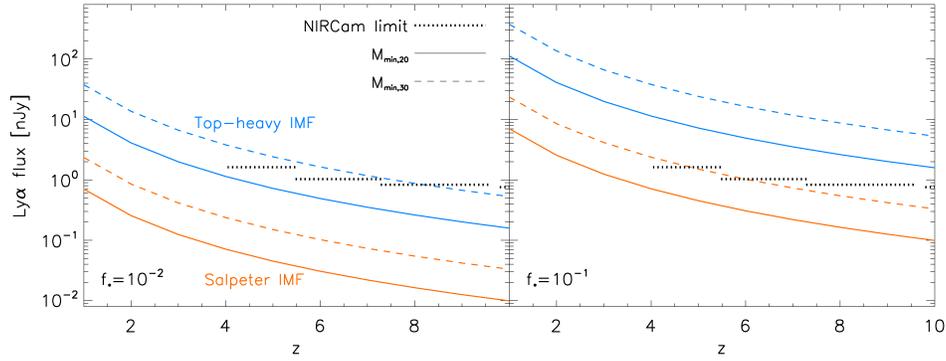}
  \caption{The monochromatic Ly$\alpha$ flux of Pop~III clusters as would be observed by the NIRCam instrument in the JWST DWS, 
for different choices of the IMF, star formation efficiency $f_{\rm *}$, and minimum halo mass $M_{\rm min}$ for star formation, as described in \protect\cite{JLJ2010}.  
The {\it black dotted} lines are the 3 $\sigma$ detection limits expected for the DWS, for an exposure time of 2 $\times$ 
10$^5$ seconds.  For a star formation efficiency of $f_{\rm *}$ = 10$^{-2}$ ({\it left panel}) only clusters with a top-heavy IMF may be detected, while 
for a high star formation efficiency $f_{\rm *}$ = 10$^{-1}$ ({\it right panel}) even clusters with a Salpeter IMF may be detectable out 
to $z$ $\sim$ 6.}
\end{figure}

\end{document}